\documentclass[ba]{imsart}
\pubyear{2021}
\volume{TBA}
\issue{TBA}
\firstpage{1}
\lastpage{1}

\usepackage{amsthm}
\usepackage{amsmath}
\usepackage{natbib}
\usepackage{natbib}
\usepackage[colorlinks,citecolor=blue,urlcolor=blue,filecolor=blue,backref=page]{hyperref}
\usepackage{graphicx}
\startlocaldefs
\endlocaldefs

\begin{document}

%% *** Frontmatter *** 

\begin{frontmatter}
\title{New spatial models for integrating standardized detection-nondetection and opportunistic presence-only data: application to estimating bird mortality hotspots linked to powerlines}
%\title{\support{}}
\runtitle{}

\begin{aug}
\author{\fnms{Jorge} \snm{Sicacha-Parada} \thanks{jorge.sicacha@ntnu.no}}
\author{\fnms{Diego} \snm{Pavon-Jordan} \thanks{diego.pavon-jordan@nina.no}}
\author{\fnms{Ingelin} \snm{Steinsland} \thanks{ingelin.steinsland@ntnu.no}}
\author{\fnms{Roel} \snm{May} \thanks{roel.may@nina.no}}
\author{\fnms{B\aa rd} \snm{Stokke} \thanks{bard.stokke@nina.no}}

%\author{\fnms{<firstname>} \snm{<surname>}\thanksref{}\ead[label=e1]{}}
%\and
%\author{\fnms{} \snm{}}

%\runauthor{	}

%\thankstext{<id>}{<text>}

\end{aug}

\begin{abstract}
	The constant increase in energy consumption has created the necessity of extending the network of energy transmission. Placement of powerlines represent a risk for bird population. Hence, better understanding of deaths induced by powerlines, and the factors behind them are of paramount importance to reduce the impact of powerlines. To address this concern, professional surveys and Citizen Science data are available. While the former data type is observed in small portions of the space by experts through expensive standardized sampling protocols, the latter is opportunistically collected over large extensions by citizen scientists.\\ \\
		In this paper we set up full Bayesian spatial models that 1) fusion both professional surveys and Citizen Science data and 2) explicitly account for preferential sampling that affects professional surveys data and for factors that affect the quality of Citizen Science data. The proposed models are part of the family of Latent Gaussian Models as both data types are interpreted as thinned spatial point patterns and modeled as Log-Gaussian Cox processes. The specification of these models assume the existence of a common latent spatial process underlying the observation of both data types.
		
		The proposed models are used both on simulated data and on real-data of powerline-induced death of birds in the Trøndelag county in Norway. The simulations studies clearly show increased accuracy in parameter estimates when both data types are fusioned and factors that bias their collection processes are properly accounted for. The  real data application contributes to the identification of risk factors and riskier powerlines for bird populations. The choice of model was relevant for the conclusions that could be drawn from this case study as different models estimated the association between risk of powerline-induced deaths and the amount of exposed birds differently. There is clearly association between powerline density and the risk of powerline-induced deaths.
\end{abstract}

%% ** Keywords **

\end{frontmatter}

	\section{Introduction}
	
%The perpetual human population growth comes at many costs, inter alia an ever increasing demand for energy supply. 
Energy consumption is anticipated to rise by ca. 50\% by 2050 \citep{conti2016international} and hence the global network for energy transmission must also be extended, particularly to meet the UN’s Sustainable Development Goal (SDG) 7 “\textit{universal access to affordable, reliable, and modern energy services}” \citep{UN2018}. In addition to the high expenditure that this implies \citep{bernardino2018bird}, such a development in infrastructure and land sparing will have an enormous environmental cost \citep{biasotto2018power}. Accumulating evidence shows that powerlines are an important threat for many avian species \citep{martin2011understanding}, with overhead wires fragmenting the airspace used by birds, increasing mortality risk by collision \citep{bernardino2018bird} and with masts used by many species as perching structures, causing increased death rates by electrocution \citep{hernandez2018start}. \citet{davis2002roadmap} estimated that the deaths caused by powerlines easily reach a billion birds per year. Since then, the network of powerlines has increased by 5\% annually  \citep{jenkins2010avian},
%, and despite some effective local mitigation actions (Pavón-Jordán et al. 2020, Barrientos et al. 2012), that estimate must surely be obsolete and 
and the number of deaths has most likely increased despite the success of effective local risk mitigation actions  %have been successful 
\citep{pavon2020birds,barrientos2012wire}.\\ \\%, but are expencive?. 
From the conservation point of view and in line to the SDG 9 “\textit{build resilient and sustainable infrastructure}”, SDG 11 “\textit{make cities and human settlements sustainable}”, SDG 12, “\textit{ensure sustainable consumption and production patterns}”, SDG 15 “\textit{protect, restore and promote sustainable use of terrestrial ecosystems and halt biodiversity loss}”\citep{UN2018}, it is of paramount importance to
%pinpoint the main factors involved in powerline-induced mortality and to 
gain a better understanding of under which circumstances (i.e. how, where and when) powerlines suppose a high death hazard. This can target mitigation actions and plan new power lines such that their ecological impact is reduced.
%or otherwise their impact is minimal.
\\ \\ 
Until now studies of mortality impact of powerlines have been based on
%Finding overall impacts of powerlines on birds is not a trivial task because mortality studies using 
standardised detection-nondetection data (sensu \citet{Milleretal2019}), are often carried out at specific locations and relatively small spatial scale (e.g. \citet{Bevanger1995,bevanger2001bird}) and/or focus on target species of concern (e.g. \citet{janss2001avian,lopez2011solving}). This hinders our ability to draw broader conclusions about the factors involved in this recognised human-wildlife conflict. One way to increase the range of species and habitats as well as the geographical extent represented in such datasets is to use the vast information contained in Citizen Science (CS) portals (e.g. https://ebird.org, www.artsobservasjoner.no). Some of these CS platforms allow citizen scientists to report additional information on their observations, including whether the finding was a dead animal and even the potential cause of death (e.g. electrocution, collision with powerline wires, collision with fences, roadkill). These two sources of information (standardised detection-nondetection and opportunistic presence-only data), however, are not directly comparable as they come with different inherent biases, especially regarding survey effort \citep{botella2021jointly}.\\ \\
Aware of the potential benefits the integration of multiple data types has to offer, much research has in the past years attempted to overcome the challenges of integrating more than one data source \citep{koshkina2017integrated,Pacifici2017,Milleretal2019,doi:10.1002/ecm.1372,zipkin2021addressing,wang2021combining}. Evidence of the benefits in inferential and predictive performance when multiple data types are integrated is accumulating \citep{Simmonds2020,doi:10.1002/ecm.1372}. These benefits include, inter alia, reduction in the uncertainty of the predicted variable in comparison to the results when each data type is modelled separately and increased accuracy in parameter estimation. Nevertheless, accounting for potential biases in the collection process is of paramount importance for fusion models to perform as expected \citep{Simmonds2020}.\\ \\
This paper is motivated for a case study whose aim is to determine which factors are associated to high risk of powerline-induced death of birds, and hence highlight riskier areas within the powerline network of Trøndelag, Norway. Two data types are available to address this question, data collected through professional surveys performed by the Norwegian Institute for Nature Research (NINA) and opportunistic records collected from two sources: i) Artsobservasjoner, a database of the Norwegian Biodiversity Information Center (NIBC), and ii) the Norwegian Bird Ringing Centre. Extensive research in this field show that factors such as visibility, land use, the density of the powerlines and the amount of bird that are exposed to collide with the powerlines are associated to the risk of powerline-induced deaths \citep{Bevanger2014,drewitt2008collision,martin2010bird}.\\ \\ 
Detection-nondetection data collected in professional surveys for ecological research are often analysed using site-occupancy models \citep{mackenzie2002should} and geostatistical spatial models \citep{Banerjee2015} as the data locations are fixed and defined during the sampling design. %In Section 2, we will introduce our two data types. 
For our case study, the data collected by NINA contains exact locations of where powerline-induced deaths have occurred. As a census is performed in the area around a powerline once it has been selected to be sampled, we regard these data as a thinned point pattern \citep{illian2008statistical} with thinning probability depending on the area to which the points belong. If the removal of points from the original point pattern is driven by Missing Not At Random (MNAR) mechanisms \citep{little2019statistical} that depend on our ecological process, we have thinning caused by a preferential sampling design \citep{Diggle2010}. Otherwise, the removal of points is assumed to occur randomly.\\ \\
CS data are also regarded as a thinned point pattern, but due to a MNAR mechanism \citep{little2019statistical} that depends on factors such as differences in the sampling effort, detectability, reporting effort and/or misclassification. A general flexible framework for generating CS data has been proposed by \textbf{Peprah, Sicacha-Parada, 2022}. This framework links thinning operations for point patterns \citep{illian2008statistical} with the biases in CS data and provides a novel perspective for modeling CS data. This framework relies on the idea of a shared process model for modeling data generated through MNAR mechanisms \citep{little2019statistical}. Hence, this model assumes a common latent effect that drives multiple observed data. In our case, we assume this common latent effect affects both the observed data and the missingness process. For our case study CS data is obtained from two sources, both of them share biases, such as uneven sampling effort, which can be affected by factors such as accessibility and/or land use \citep{Monsarrat2019,sicacha2021accounting}, differences in detectability, which can be explained by land use, habitat type, the size of the dead bird and/or moment of the observation \citep{dominguez2020factors}, and uneven reporting effort \citep{august2020data}.\\ \\
Hence, the aim of this paper is to introduce a modeling framework that integrates professional surveys and CS data while accounting for the biases in the collection of each of the data types as suggested in \citet{Simmonds2020}. As a natural result of the modeling framework proposed, we also expect to highlight riskier areas for powerline-induced deaths. This framework extends the state of the art of data integration models as it simultaneously models CS data and their biases and professional surveys data and their sampling process, which might be preferential \citep{Diggle2010}. This modeling framework is specified as a group of Bayesian models that depend on shared spatial random effects and lies within the class Latent Gaussian Models (LGMs, \citet{Rue2009}). As these models belong to the family of LGMs, they are suitable for being fitted using both the Integrated Nested Laplace Approximation (INLA) \citep{Rue2009} and the Stochastic Partial Differential Equation (SPDE) \citep{Lindgren2011} which offer efficient approximation of both the posterior distribution of the parameters of the models and the Gaussian Random Fields (GRF) involved in the specification of the models \citep{10.1093/biomet/asv064}. This is a flexible framework for integrating the two available data types and accounting for multiple sources of bias in both professional surveys and CS data. Hence, the models we present can be applied directly in assessments not only of the impacts of powerlines on animal mortality (our case study), but also in those of other human infrastructures such as roads and windfarms. Such assessments are critical in times where the rapid increase of new projects linked to renewable energy development are adding mortality to that of the traditional roadkills \citep{barrientos2021lost} and powerlines \citep{bernardino2018bird} and are facing strong public rejection \citep{serrano2020renewables}.\\ \\
We show the performance of the models we propose for both simulated data and our case study. Through the simulation study we show the relevance, potential benefits for parameter estimation and challenges of integrating both data types. In the case study, despite not knowing the ground truth, we expect to find riskier areas for powerline-induced deaths, which factors determine this risk, as well as comparing each of the models proposed.\\ \\
This paper is organized as follows. In Section 2 we present the two available data types for our case study as well as details of their collection process that are relevant for the specification of our modeling framework. This framework is introduced in Section 3, technical details regarding the models for integrating the two data types and accounting for their biases are also presented in this section. In Section 4, the simulation studies to assess the properties of our models as well as the necessity of accounting for biases and of integrating both data types are presented. Section 5 contains the analysis of powerline-induced death of birds and the results. Finally, in Section 6 we discuss the proposed framework and propose future extensions of it.

\section{Data and case study: Death of birds caused by powerlines in Trøndelag, Norway}

\begin{figure}[H]	
	\center
	\includegraphics[width=0.6\textwidth]{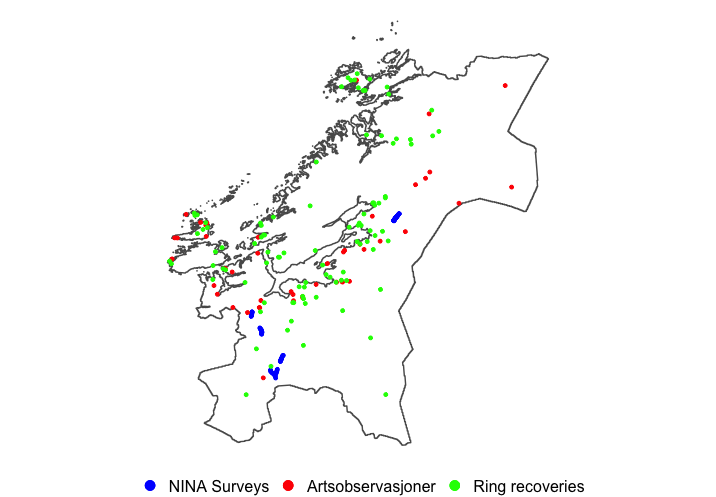} 
	\caption{\textit{Data sources considered for studying the risk of powerline-induced bird mortality and their spatial distribution. In blue: Spatial distribution of observations collected by NINA. In red: Opportunistic records reported by citizen scientists in Artsobservasjoner. In green: Opportunistic records reported through the Norwegian Bird Ringing Centre.}}
	\label{datatypes}
\end{figure}

Here we use two different datasets on bird casualties due to powerlines. First, we use standardised data from professional surveys conducted by the Norwegian Institute for Nature Research (NINA) aiming at finding all bird carcasses under a specific section of a powerline \citep{Bevanger2014}. Second, we use two opportunistic presence-only CS data: 1) records found at the Norwegian CS portal www.artsobservasjoner.no and reported as “dead bird by a powerline” since 2016 and 2)  Records of dead birds reported to the Norwegian Bird Ringing Centre.	Figure \ref{datatypes} displays the occurrence of both types of data. Since the detection of birds can occur up to about 100m away from a powerline, and for computational convenience, the spatial domain of the case study is a buffered version (100m on each side) of the networks of powerlines. In this study we use data from  Trøndelag, in central Norway. The large variation in environmental conditions of this county and the existence of areas with high abundance of birds \citep{sicacha2022spatial} makes this region suitable for our case study. 
	
	\subsection{Professional surveys of bird mortality}

The data of professional surveys on bird mortality caused by powerlines were collected within the OPTIPOL project \citep{Bevanger2014}. Carcasses were searched for using a trained dog (wachtel) under a 7.1 km-long section of a high-voltage powerline (300 kV transmission line) during 2011-2014. Every year, the same dog patrolled the same section following the same protocol - crisscrossing under the powerline in the clearcut area (see \citet{Bevanger2014} for further details and study design). Found carcasses were removed to avoid double-counting. For this study we used 147 observations of carcasses collected at point level as the exact geographic location is available for each observation.

%But how are the data?! Point or areal. Or we use areal for each poweline (segment)?!

%In principle, we assume these data are preferentially sampled as we assume expert prior knowledge determines which powerlines are visited and sampled. Once a powerline is visited, thorough searches are performed with a trained dog. Hence, we also assume there are no biases induced by the sampling protocols so that these data form a thinned point pattern, only affected by the probability of selecting each powerline. 

	\subsection{Opportunistic records of bird mortality}
	
 We retrieved opportunistic (presence-only) records from two sources. First, Artsobservasjoner, a database of the Norwegian Biodiversity Information Center (NBIC;www.artsobservasjoner.no) where everyone, volunteers and professionals, can report the occurrence of any species alongside the location, date/time and additional information that the observer deems important to be linked to the observation. For example, observers can report if the observed occurrence was dead and the cause of death if it is straightforward (e.g. electrocuted bird, broken wing due to collision). The second source of opportunistic records is the Norwegian Bird Ringing Centre, run by the Natural History Museum  in Stavanger (https://www.museumstavanger.no/en/forskning/den-norske-ringmerkingssentralen-1), whose database registers more than 8 million entries, including ringing data (i.e. tagging birds with metal rings with a unique identifier) and ring-recovery data. The recovery data (i.e. when a bird with a metal ring is found dead and reported to the national ringing office) allows us to, inter alia, gather information on location and causes of death (provided that the observer reports the cause of death, which may not always be the case). As precise geographic information is available for each report of this data type are available, these data are handled as point-level data.\\ \\
In total, for our case-study in Trøndelag County, we used 98 observations from ring recoveries (dead birds found with a metal ring and were reported to the ringing center) and 46 observations of dead birds killed by a powerline from the CS portal Arts Observasjoner (n = 144). 
    
   \subsection{Explanatory variables}
Two groups of explanatory variables are required to fit the models proposed in the upcoming sections. The first group of candidate variables aim to explain the ecological process underlying the powerline-induced deaths and the second group explain the sampling process of both the professional surveys and the opportunistic records.\\ \\
In the group of candidate variables for explaining the ecological process, we include elevation (Digital Elevation Model; DEM), mean temperature and precipitation (Norwegian Meteorological Institute), bird abundance estimated from the Norwegian common bird monitoring scheme (https://tov-e.nina.no; see also \citet{sicacha2022spatial}),  powerline density (The Norwegian Water Resources and Energy Directorate; NVE) cloud cover (https://www.earthenv.org/cloud) and land cover (AR50, https://www.nibio.no/tema/jord/arealressurser/ar50). All covariate information was rasterized to a scale of 1 x 1 kilometers. The second group of variables explain the sampling process of the professional surveys and the CS data. The candidate covariates to explain the sampling process include distance to the closest (tertiary) road (road network obtained from OpenStreetMap https://www.openstreetmap.org), as well as the distance to the nearest water body (where citizen scientists frequently go for birdwatching due to high bird abundance). To explain the sampling process that yield the observed professional surveys data, we use elevation gradient.
\subsection{Exploratory analysis}
\begin{figure}[H]	
	\center
	\includegraphics[width=0.8\textwidth]{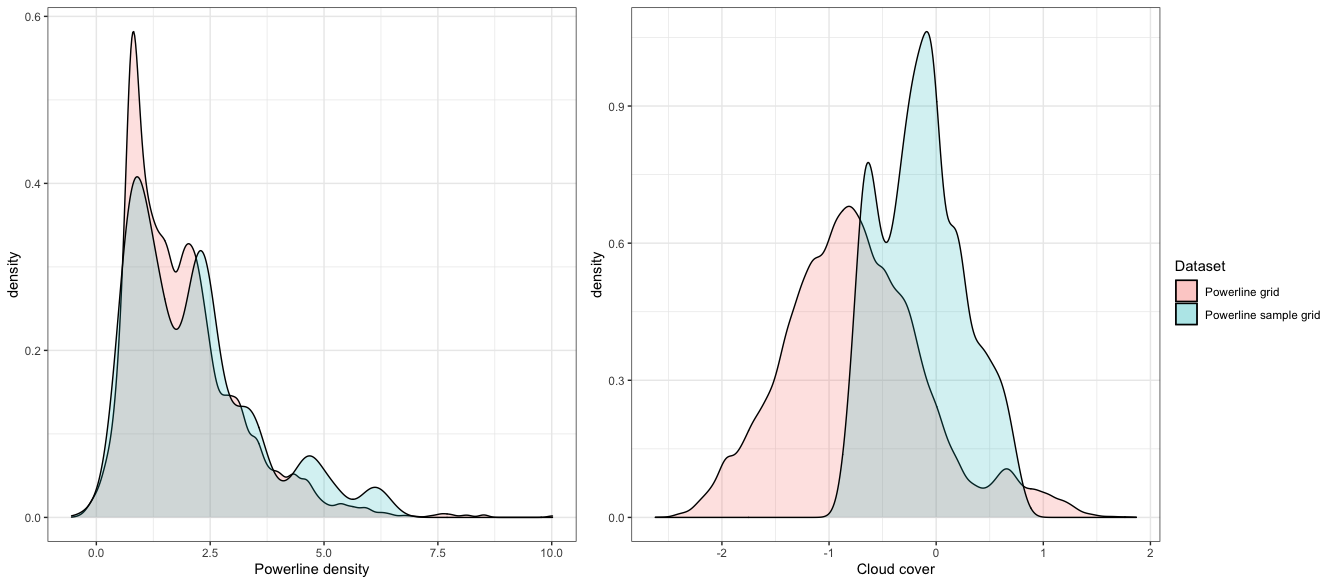} 
	\caption{\textit{Density plot for powerline density and cloud cover for the two datasets previously defined. In red, the density plot for the grid defined over the whole network of powerlines. In blue, the density plot defined for the grid of points defined for the sampled powerlines.}}
	\label{prefsamptest}
\end{figure}

\begin{figure}[H]	
	\center
	\includegraphics[width=0.8\textwidth]{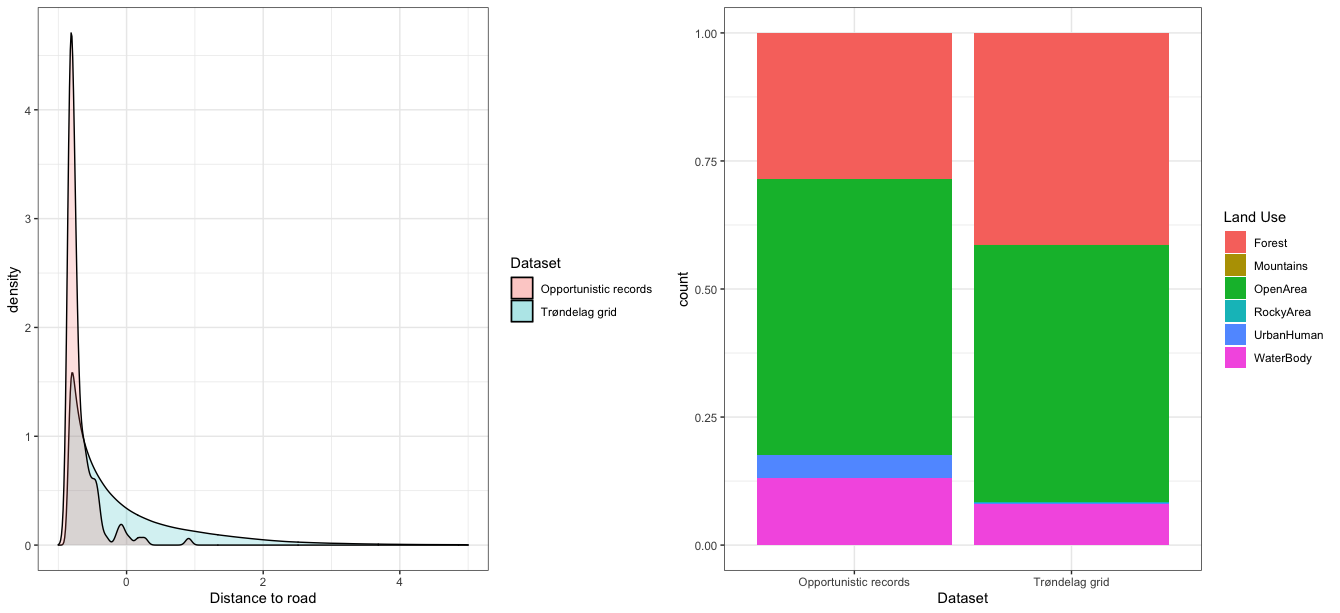} 
	\caption{\textit{Density plow for distance to (tertiary) roads and partition of land uses for the two datasets previously defined. On the left panel: in blue, the density plot for the grid defined over the whole county of Trøndelag; in red, the density plot defined for the locations with opportunistic records. On the right panel: left bar: partition of land use in the locations with opportunistic records; right bar: partition of land use over the whole county of Trøndelag}}
	\label{CSbiastest}
\end{figure}

Despite the standard sampling effort made in NINA’s professional surveys, these data have some limitations. First, their spatial coverage is small as carrying out these surveys is time-consuming and expensive. Second, the selection of the survey sites (powerlines) is not completely random \citep{Bevanger2014} as expert knowledge is used to determine which powerlines should be visited. Now we explore the sampled locations during the CS projects and determine whether or not there is indication of preferential sampling in these data. To do it, we make two datasets, one with a 100mx100m grid of points along a 100-meter buffer of the network of powerlines in Trøndelag and the other one with a 100mx100m grid of points defined over the powerlines that have been visited by NINA. Our comparison focuses on two covariates, powerline density and cloud cover. The results are presented in Figure \ref{prefsamptest}. Note that both covariates are standardized based on their values over Trøndelag.
Even though both densities seem similar for the powerline density, we notice that denser powerlines are more frequent among the sampling locations chosen by the experts. On the other hand, the right panel of Figure \ref{prefsamptest} shows that those powerlines selected by the experts are located where cloudcover is higher that usual along the network of powerlines. It is known that both powerline density and visibility (proxied by cloud cover) are factors associated with powerline-induced mortality \citep{drewitt2008collision}. Hence, there is apparently indication of preferential sampling in the sampling design performed by experts at NINA.\\ \\
Both Artsobservasjoner data and ring recoveries can be regarded as CS data. Both of them share biases, such as uneven sampling effort, differences in detectability and uneven reporting effort. The latter might be regarded as an important source of bias as it is more common for rare occurrences, as in the case of dead birds, which might not be as convenient to report as the occurrence of alive individuals. Now, we explore the available CS data and determine if there is indication of a sampling design affected by factors such as accessibility or land use. We define again two datasets in this case. The first one, a 100mx100m grid of points defined over Trøndelag and the second one, the locations where powerline-induced bird death have occurred. The results are displayed in Figure \ref{CSbiastest}.
On the left panel of Figure \ref{CSbiastest} we notice that the locations where citizen scientists have reported occurrences of dead birds are considerably closer to tertiary roads than all the other locations in the study area. Regarding the land use, we see from the right panel of Figure \ref{CSbiastest} that the percentage of reports of dead birds that occur in open areas is higher than the percentage of open areas in Trøndelag. On the other hand, the percentage of reports made in forest areas is much smaller than the actual portion of Trøndelag that is covered by forest. As it is known that accessibility has no association with the risk of powerline-induced mortality, there seems to be an indication of bias due to accessibility in the sampling process of CS data, as previously mentioned in \citet{fithian2015bias,Monsarrat2019,sicacha2021accounting}. Unlike accessibility, land use might be associated with differences in the risk of death due to powerlines. However, it is a factor that is also associated with higher or lower detectability. Therefore, in the upcoming sections we will examine whether or not land use should be considered as a factor linked to differences in risk of powerline-induced death and differences in detectability.

	\section{Models}

%I'm not sure about the model, so below is my best try:

%\begin{itemize}
 %   \item True log intensity: $\log(\lambda_{true} (\textbf{s})) = X(\textbf{s})\beta + \omega_1(\textbf{s})$.
%    \item True process a realization of this (NHGP?)
%    \item Both CS and PA(?) observation processes of form $NHGP (\lambda_{true}(s) b(s) \delta (s))$ (or is it Bernoulli for PA? Connection to $\labmda_{\true}$ through logit?)
%    \item For CS:
 %   \begin{itemize}
 %       \item $log (b_{CS}(s)) = Z(s) \gamma + \omega_2(s)$
%        \item $\delta$ two different
%    \end{itemize}
%    \item For PA
%    \begin{itemize}
%        \item $\log (b_{PA}(s)) = W(s) \alpha + \zeta_1 \omega_1(s)$. 
  %      \item $\delta(s)=1$ (observe and report all)
 %   \end{itemize}
 %   \item Can see the observation models as P(is a electrified bird) P(go looking $|$ is) P(report $|$ is and go)
%\end{itemize}

The  generating process of occurrence of dead birds related to the network of powerlines in Trøndelag is the target model of this work. We model occurrence of dead birds along the network of powerlines $P \subset \mathbb{R}^2$ as a point pattern with intensity $\lambda_{true}(\textbf{s})$. This intensity is further modeled as a sum of a linear combination of known environmental variables $X(\textbf{s})$ as, for example, powerline density and abundance of birds, and a spatial Gaussian Random Field, $\omega_1(\textbf{s})$:
\begin{equation}
\log(\lambda_{true}(\textbf{s})) = X(\textbf{s})\beta + \omega_1(\textbf{s}) 
\label{truepp}
\end{equation}
This true intensity of birds dead due to powerlines and its components are the quantities of interest. In our case study we have two data sources available, the citizen science (CS) data and the professional data as described in Section 2. Each data type has two different observation processes which we propose models for below. 

\subsection{Modelling data process of professional  surveys}
There are two aspects we want to include in our model of the professional surveys: 1) sampling locations might be chosen preferentially, and 2) the sampling effort is very high and we assume that the process is perfectly sampled. Therefore, we interpret the observed data by experts as a realization of the preferential sampling process described by \citet{Diggle2010}. 
%The network of powerlines $P$ is partitioned in (buffered) $m$ power lines. 
Each of the $m$ buffered powerlines (see Section 2) has a probability $\phi_i(\textbf{s}); i=\{1,\ldots,m\}$ of being sampled. If the selection of the lines that are visited is completely random we assume $\phi_1(\textbf{s}) = \ldots = \phi_m(\textbf{s})$. However, as in our case study, experts usually have prior knowledge on where they can find what they want to investigate, we assume the probabilities in $\boldsymbol \Phi= [\phi_1(\textbf{s}),\ldots,\phi_m(\textbf{s})]^T$ are stochastically dependent on the latent process $\omega_1(\textbf{s})$ (see Eq. \ref{truepp}) and eventually on some of the covariates in $X(\textbf{s})$. In addition to $\omega_1(\textbf{s})$, the probabilities in $\boldsymbol \Phi$ are explained by known variables $Z(\textbf{s})$, which might include some of the covariates in $X(\textbf{s})$,  as follows:
\begin{equation}
logit(\phi_i(P_i)) = Z(P_i)\gamma + \zeta \omega_1(P_i)
    \label{prefsampling}
\end{equation}
The coefficient $\zeta$ is relevant for this specification as it determines the extent of the dependence between the sampling selection of power lines to be sampled and the true ecological process that defines hotspots for bird collision or electrocution \citep{doi:10.1002/ecm.1372}.\\ \\
Once the powerlines to be sampled have been selected, we assume, as stated at the beginning of this subsection, that within each selected area, a complete census with perfect detection of the dead birds is performed. Hence, we observe a point pattern $\textbf{P}_{PA}$ with intensity $\lambda_{PA}(\textbf{s}) = \lambda_{true}(\textbf{s}) \phi_k (\textbf{s})$, with $k$ the powerline where location $\textbf{s}$ belongs. Figure S.1. explains graphically the observation process performed in professional surveys.\\ \\
It is worth noting that as we work with a buffered version of the network $P$, the model in Eq. (\ref{prefsampling}) is an areal data model \citep{Banerjee2015} and accounting for preferential sampling while avoiding identifiability issues implies modelling jointly $\lambda_{PA}(\textbf{s})$ and the selection probabilities $\phi_i(\textbf{s})$. More details about this part of this modelling framework are provided in Section 3.1.1.

\subsubsection{Modelling preferential sampling}

As previously stated, modelling preferential sampling for our case study implies the integration of a point process and areal data. Both data types depend on the Gaussian Random Field $\omega_1(\textbf{s})$. However, while a continuous version of the GRF generates the point process, a discretized version of it produces the areal data. Integration of two or more types of spatial data has been previously approached by \citet{roksvaag2020estimation} and \citet{wang2021combining} with applications to hydrology and epidemiology, respectively. In general, a GRF at the area $\textbf{A}$ is expressed and approximated as:
\begin{equation}
    \omega_1(\textbf{A}) = \frac{1}{|\textbf{A}|} \int_{u \in \textbf{A}} \omega_1(u) du \approx \frac{1}{H} \sum_{\textbf{s} \in \textbf{A}} \omega_1(\textbf{s})
\end{equation}
with $\textbf{s}=\{s_1,\ldots,s_H\}$ a set of sampling points in $\textbf{A}$. However, as $\omega_1(P_i)$ is a nonlinear function of $\phi_i(P_i)$  in Eq. (\ref{prefsampling}), we express and approximate $\omega_1(\textbf{A})$ for our case study as:
\begin{equation}
    \omega_1(\textbf{A}) = logit \left( \frac{1}{|\textbf{A}|} \int_{u \in \textbf{A}} \frac{\exp(\omega_1(u))}{1+\exp(\omega_1(u))} du \right) \approx logit \left( \frac{1}{H} \sum_{\textbf{s} \in \textbf{A}} \frac{\exp(\omega_1(\textbf{s}))}{1+\exp(\omega_1(\textbf{s}))} \right)
\end{equation}
as presented in \citet{wang2021combining} in order to avoid ecological bias \citep{greenland1992divergent}.

 \subsection{CS data process model}
We assume CS data is the result of a thinning process of the true point pattern of dead birds with three components: sampling effort, detectabilty and reporting effort. They can be sources of bias due to uneven sampling effort understood as differences in the probability of location $\textbf{s}$ being sampled by citizen scientists, due to differences in detectability according to the location of the occurrence of the event and due to differences in reporting probability amongst  the observers. 
%Each thinning stage corresponds to a source of bias. 
%In our case, these sources of bias include uneven sampling effort understood as differences in the probability of location $\textbf{s}$ being sampled by citizen scientists. This probability may depend on environmental covariates at $\textbf{s}$. Another source of bias that drives the observed point pattern is caused by differences in reporting probability amongst all the observers. 
Based on this, we assume CS data follow a point processs model, to be specific a log Gaussian Cox process (LGCP), $\mathbf{P_{CS}}$ with intensity $\lambda_{CS}(\textbf{s})$
given by:
\begin{equation}
\lambda_{CS}(\textbf{s}) = \lambda_{true}(\textbf{s}) \cdot \tau(\textbf{s}) \cdot \psi(\textbf{s}) \cdot \delta(\textbf{s})
\label{obspp}
\end{equation}
where 
$\lambda_{true}(\textbf{s})$ is the intensity of the  true occurrences of dead birds due to powerlines, 
$\tau(\textbf{s})$ is the %probability of retaining a true occurrences 
probability of location $\textbf{s}$ being part of the locations sampled by citizen scientists, $\psi(\textbf{s})$ is the probability of detecting an occurrence at location $\textbf{s}$ given that it has been visited and  $\delta(\textbf{s})$ is the probability of reporting an occurrence at $\textbf{s}$ given that the location has been visited by a citizen scientist and an occurrence has been detected.\\ \\ %As pointed out previously, several sources of bias affect the knowledge we can get from CS reports about death of birds due to power lines. In our particular problem 
 We focus on accounting for three sources of bias: uneven sampling effort , detectability and reporting effort. 
 However, this modeling framework is flexible enough as to account for any other source of bias that might need to be accounted for.
 
 \subsubsection{Sampling effort}
 The sampling process of CS data is not standardized \citep{doi:10.1111/2041-210X.12254} and is often biased towards locations with higher accessibility, or locations where observers expect to find more occurrences, i.e. locations that are preferentially sampled, \citep{fithian2015bias,Monsarrat2019}. Hence, we propose modeling the term $\tau(\textbf{s})$ as:
 \begin{equation}
 logit(\tau(\textbf{s})) = Z(\textbf{s})\alpha + \omega_2(\textbf{s})
 \label{CSsamp}
 \end{equation}
 where $Z(\textbf{s})$ are covariates that aim to explain which locations are most likely visited by citizen scientists and $\omega_2(\textbf{s})$ is a spatial random effect.
 \subsubsection{Differences in detectability}
 In many real-life scenarios the detection of an event of interest, (a dead bird in our case) can not be assumed constant. For our case-study, a relevant factor that affects whether or not a dead bird can be detected is the land use. As pointed out in previous studies \citep{dominguez2020factors}, habitats such as open areas allow for easier detection of such events. Hence, the differences in detectability are regarded as a second thinning factor for the actual point pattern of dead birds. We define the probability of detecting the occurrence of a dead bird at location $\textbf{s}$ given that it has been visited as:
  \begin{equation}
 logit(\psi(\textbf{s})) = W(\textbf{s})\nu 
 \label{CSdet}
 \end{equation}
 with $W(\textbf{s})$ the covariates that explain the factors that affect detectability, in our case land use or habitat type.
 \subsubsection{Reporting effort}
 
Estimating the reporting effort represented in $\delta(\textbf{s})$ requires information that is almost always unavailable. However, we here propose two ways of accounting for the noise that the uneven reporting effort generates on the observed point pattern of reports of dead birds.\\ \\
 \textbf{Simple report effort model\\ \\}
 This proposal assumes no structure in the random effect that drives the differences in reporting error. Hence, $\delta(\textbf{s})$ is assumed to depend on a hyperparameter $\theta$. The relation between $\delta(\textbf{s})$ and $\theta$ is given by:
 \begin{equation}
 \delta(\textbf{s}) = \frac{\exp(\theta)}{1+\exp(\theta)}
 \label{delta1}
 \end{equation}                                         
 where $\theta$ has a normal prior $\pi(\theta) = N(0,1)$.\\ \\                                        
 \textbf{Observer-specific report effort model\\ \\}
 This is a more complex approach as we assume the probability of reporting associated to each occurrence depends on its location and the observers whose citizen science activity occurs around it (i.e. how active an observer is around the location of the occurrence based on her/his other reported observations). In this case we express $\delta(\textbf{s})$, the probability of reporting an occurrence at location $s$ given that it was visited by a citizen scientist and an occurrence was detected as:
 \begin{equation}
 \delta(\textbf{s}) = \sum_{j \in \{obs\}} \psi_j(\textbf{s}) \kappa_j(\textbf{s})
 \label{delta2}
 \end{equation}
 where $\psi_j(\textbf{s})$ is the probability that the observer of an occurrence that was sampled is observer $j$ and $\kappa_j(\textbf{s})$ is the probability that this occurrence that observer $j$ has detected becomes finally reported. The expression in (\ref{delta2}) can be interpreted as a weighted average of the probabilities that each observer reported the occurrence of a dead bird once they saw one. The weights $\psi_j(\textbf{s})$ depend on characteristics of the observers such as the distance to $\textbf{s}$, or how active the observer (i.e. number of observations in the CS portal) is. In this modeling framework, $\psi_j(\textbf{s})$ is taken as a deterministic input while the aim of the model is to estimate $\kappa_j(\textbf{s})$. A broader explanation of how $\psi_j$ could be estimated is available in the Supplementary Information.
 
 \subsection{Prior specification}
 
The spatial GRFs $\omega_1(\textbf{s})$ and $\omega_2(\textbf{s})$ in Eq (\ref{truepp}) and Eq (\ref{CSsamp}) are assumed to follow a Matérn covariance function given by:
 	\begin{equation}
	\frac{\sigma^2}{\Gamma(\nu)2^{\nu-1}}(\kappa \|s_i-s_j\|)^{\nu} K_{\nu} (\kappa\|s_i-s_j\|)
	\label{MaternCov}
	\end{equation}
	with $\|s_i-s_j\|$ the Euclidean distance between two locations $s_i$, $s_j \in D$. $\sigma^2$ stands for the marginal variance, and $K_{\nu}$ represents the modified Bessel function of the second kind and order $\nu >0$. $\nu$ is the parameter that determines the degree of smoothness of the process, while $\kappa>0$ is a scaling parameter. The parameter $\nu$ is fixed to be 1. The spatial range $\rho$ is expressed as $\rho=\sqrt{8}/\kappa$. The prior distribution of the parameters $\rho$ and $\sigma$ are specified by making use of Penalized Complexity (PC) priors, \citep{doi:10.1080/01621459.2017.1415907}. The parameter vectors $\boldsymbol \beta$, $\boldsymbol \gamma$, $\boldsymbol \alpha$ and $\boldsymbol \nu$ have Normal prior distribution with mean 0 and precision 0.01.
 \subsection{Fitting the models: inlabru R-package}
 As our modeling framework lies within the framework of the Latent Gaussian Models (LGMs; \citet{Rue2009}), our models can be conveniently fitted using the INLA-SPDE approach \citep{Rue2009,Lindgren2011}. INLA produces fast, reliable inference as it aims to produce a numerical approximation of the marginal posterior distribution of the parameters and hyperparameters of the model. The SPDE approach is based on the solution of a SPDE which can be approximated through a basis function representation defined on a discretization of the spatial domain. \citet{10.1093/biomet/asv064} propose using the SPDE approach based on a tesellation of the space to efficiently approximate the likelihood of a LGCP.\\ \\
This modeling framework incorporates nonlinear terms in the linear predictor in order to account for the biases in CS data. An alternative to deal with these non-linearities is to iteratively linearize them. This is done using the inlabru R-package \citep{bachl2019inlabru}, which makes an approximation by linearizing the non-linear function at a so-called linearization point and then approximating the posterior distribution through a Taylor series approximation of second order at this point. The fixed point iteration method \citep{burden2015numerical} is used to find the linearization point. Additional details and examples are available at \textit{https://inlabru-org.github.io/inlabru/articles/method.html}.

\section{Simulation Studies}
In order to explore the importance of accounting for the sampling process of CS data as well as the uneven reporting effort between citizen scientists, we conduct a simulation study based on the Trøndelag case study with the powerlines of the region and some of the explanatory variables defined in Section 2.3. We set up for simulation scenarios representing different selection schemes of the powerlines visited by the experts (random or preferential sampling) and different willingness (low or high) to report dead birds once they have been detected by citizen scientists. For each of these scenarios 100 datasets are generated and fitted to a group of models that range from models that use each of the data sources separately to models that integrate professional surveys and opportunistic records while account for some biases in the collection of both data types.

\subsection{Datasets simulation models}
\subsubsection{Simulation model of the true occurrence}
To start our simulation study, we generated 100 spatial point patterns with intensity $\lambda_{true}(\textbf{s})$ that represents the true risk intensity as:
\begin{equation}
    \log(\lambda_{true}(\textbf{s})) = -2 + 0.75\textit{ CLOUDCOVER}(\textbf{s}) + \omega_1(\textbf{s})
    \label{sim:truepp}
\end{equation}
where represents each of the 100 realizations of the GRF $\omega_1(\textbf{s})$ with range $\rho_1=1$ and variance $\sigma_1^2=0.3$. 
\subsubsection{Citizen science simulation models}
The points generated following the specification in Equation (\ref{sim:truepp}) were thinned with a thinning probability that depends on the sampling process of citizen scientists (see Sec. 3). In this case the sampling process of citizen scientists is represented through a point process with intensity $\lambda_{CS}(\textbf{s})$  specified as:
\begin{equation}
    \log(\lambda_{CS}(\textbf{s})) = -4 -2\textit{ DISTANCE}(\textbf{s}) + \omega_2(\textbf{s})
    \label{sim:cssamplingpp}
\end{equation}
where \textit{DISTANCE} represents the distance to the closest tertiary road and $\omega_2(\textbf{s})$ a GRF with range $\rho_2=100$ and variance $\sigma_2^2=1.3$. As the coefficient of the covariate \textit{DISTANCE} is  negative, those occurrences located at more accessible locations for citizen scientists have higher probability of being retained. The next stage of thinning depended on the detection probability of an occurrence of a dead bird based on factors such as land use, which affects how likely a dead bird is detected by an observer that has visited a location with an occurrence. The probability  of retaining an occurrence, $\psi(\textbf{s})$, was expressed as:
\begin{equation}
    \text{logit}(\psi(\textbf{s}))= 1 - 2\textit{MOUNTAIN}(\textbf{s})+1.2\textit{OPEN}(\textbf{s}) + 1.4\textit{ROCKY}(\textbf{s})+1.8\textit{URBAN}(\textbf{s})-3\textit{WATER}(\textbf{s})
\end{equation}
Each of the covariates in this model are indicator variables for each land use as defined in Section 2.3. The last stage of thinning depended on the probability of reporting the occurrence of a dead bird given that it has been detected and an observer has reached the place where the dead body was located. 10 observers were assumed as the population of citizen scientists and $\phi_j(\textbf{s})$, as defined in Section 3.2.3, was modeled as:
\begin{equation}
    \text{logit}(\phi_j(\textbf{s})) = 10 - 0.3DISTTOOBS_j(\textbf{s}) + \zeta ACTLEVEL_j
\end{equation}
with $DISTTOOBS_j(\textbf{s})$ the distance from the centroid of the citizen science activity of observer $j=\{1,\ldots,10\}$ to location $\textbf{s}$ and $ACTLEVEL_j$ a categorical ordinal variable with the activity level (in scale 1-5) of each observer. The vector $\zeta = [0,0.5,1,1.5,2]$ determines the magnitude of the relation between each activity level and $\phi_j(\textbf{s})$. Two definitions were given for $\kappa_j(\textbf{s})$ depending of the willingness to report of the population of citizen scientists. Low willingness to report was assumed when the values of $\kappa_j(\textbf{s})$ were generated using a $Beta(2,5)$ distribution (i.e. mean $\approx 0.29$) and high willingness when a $Beta(5,1.5)$ distribution (i.e. mean $\approx 0.77$) was used. A complete summary of the generation of each dataset is displayed in Figure S.1. in the supplementary information.
\subsubsection{Professional surveys simulation models}
These data were simulated to resemble the professional surveys carried out by NINA. For each simulated dataset we assume that the buffer around a powerline, $P_i$, is sampled with probability $\phi_i(P_i)$. This probability is assumed the same ($\phi_i(P_i)=p$) for  all the buffers when random sampling is assumed and when preferential sampling is assumed, it is given by:
\begin{equation}
    \text{logit}(\phi_i(P_i))=-2.5 -1.5 \textit{ELEVGRADIENT}(\textbf{s}) + 3.5\textit{CLOUDCOVER}(\textbf{s})
    \label{simu:prefsmodel}
\end{equation}
In this case we assume $\omega_1(\textbf{s})$ is not part  of the preferential sampling model in order to reduce the computational complexity of the simulations. Once a segment has been selected, all the occurrences within the buffer are assumed to be observed, detected and reported as the sampling effort in these surveys is high since the sampling is performed with trained dogs. A graphical summary of this sampling is presented in Figure S1

\subsection{Simulation scenarios}
\begin{figure}[H]
	\center
	\includegraphics[width=0.8\textwidth]{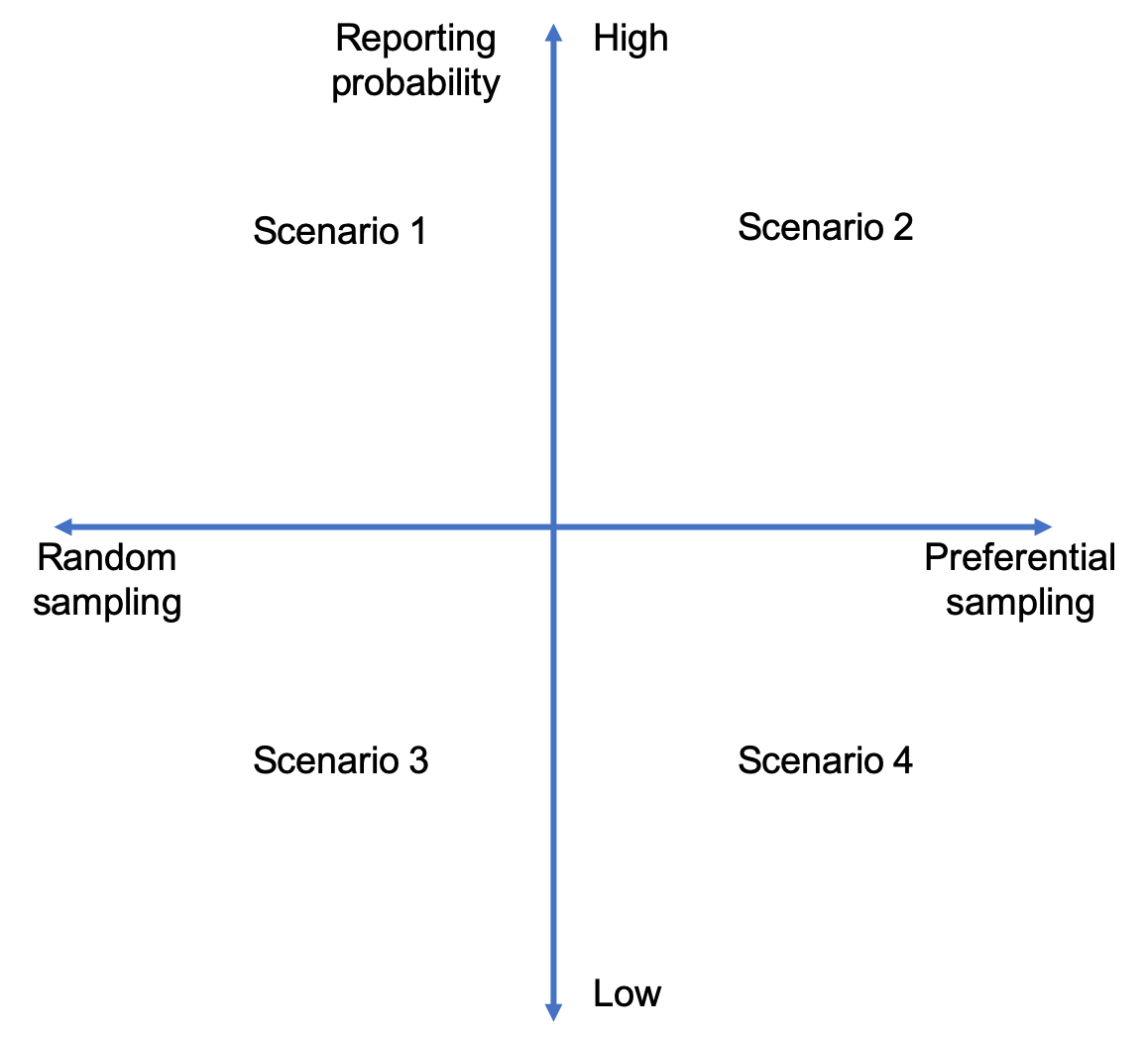} 
	\caption{Scenarios for the simulation studies of Section 4.}
	\label{fig:simuscenarios}
\end{figure}

For each simulation scenario 100 observed datasets were generated by thinning the 100 point patterns generated in Section according to Equation (\ref{sim:truepp}). The way these point patterns are thinned define the four simulation scenarios 
%depending on the willingness (low or high) of citizen scientists to report an occurrence of a dead bird and on the selection of the powerlines sampled by the experts (random or preferential sampling). A summary of these scenarios is 
displayed in Figure \ref{fig:simuscenarios}.
Two factors define the four simulation scenarios: i) whether the probability of reporting a detected occurrence is high or low (see Section 4.1.2), and ii) if the selection of the powerlines in the professional surveys is completely random or preferential (see Section 4.1.3). In Scenarios 1 and 2 we assume the probabilities of reporting occurrences that were detected by the observers are high, but in Scenario 1 the sampling of the powerlines is assumed completely random whereas in Scenario 2 the sampling of the powerlines is preferential as in Equation (\ref{simu:prefsmodel}). A similar relation holds between Scenarios 3 and 4. However, the probability of reporting for these scenarios are assumed low.
\subsection{Model comparison}
We want to compare the results of fitting different models (i.e. models that use only one data type and models that fusion both data types) for the datasets generated in the different simulation scenarios previously described. We defined the eight models in Table \ref{modcomparison} and fitted them for each of the 400 generated datasets (100 true occurrence point patterns $\times$ 4 simulation scenarios). Since models 7 and 8 are computationally expensive and specifically deviced to for scenarios with low reporting probability (scenarios 3 and 4), they were not fitted in Scenarios 1 and 2.
% Table generated by Excel2LaTeX from sheet 'Sheet5'
\begin{table}[H]
  \centering
  \caption{Description of the models fitted in the simulation study. In the data sources column: PS stands for Professional Surveys and CS  for Citizen Science.}
    \begin{tabular}{ccc}
    \toprule
    Model & Data sources & Description \\
    \midrule
    1     & PS    & \begin{tabular}{@{}c@{}}Only data from professional surveys\\ (Equation (\ref{truepp}))\end{tabular}  \\
    2     & PS    & \begin{tabular}{@{}c@{}}Only data from professional surveys accounting for preferential sampling\\ (Equations (\ref{truepp}) and (\ref{prefsampling}))\end{tabular}  \\
    3     & CS    & \begin{tabular}{@{}c@{}} Only data with locations from CS reports \\(Equation (\ref{truepp})) \end{tabular} \\
    4     & CS    &\begin{tabular}{@{}c@{}}  Same as model 3, but accounting for sampling process of CS\\ (Equation (\ref{obspp}) with $\delta(\textbf{s}$)=1) \end{tabular} \\
    5     & CS    &\begin{tabular}{@{}c@{}}  Same as model 4, but also accounting for the detection process\\ (Equation (\ref{obspp}) with $\delta(\textbf{s}$)=1) \end{tabular} \\
    6     & PS + CS & \begin{tabular}{@{}c@{}} Integration of models 2 and 5 accounting for sampling and detection process\\ of CS and preferential sampling.\\ (Equations (\ref{truepp}), (\ref{prefsampling}), and (\ref{obspp}) with $\delta(\textbf{s})$=1) \end{tabular} \\
    7     & PS + CS & \begin{tabular}{@{}c@{}} Integration of models 2 and 5 accounting for sampling, detection and\\ reporting process of CS and preferential sampling.\\ (Equations (\ref{truepp}), (\ref{prefsampling}), and (\ref{obspp}) with $\delta(\textbf{s})$ as in Equation (\ref{delta1})) \end{tabular} \\
    8     & PS + CS & \begin{tabular}{@{}c@{}} Integration of models 2 and 5 accounting for sampling, detection and\\ reporting process of CS and preferential sampling.\\ (Equations (\ref{truepp}), (\ref{prefsampling}), and (\ref{obspp}) with $\delta(\textbf{s})$ as in Equation (\ref{delta2})) \end{tabular} \\
    
    %5     &   PA + PO & \begin{tabular}{@{}c@{}}Integration of PA and PO accounting for sampling and reporting process of CS: Option 1 \\(Equations (\ref{obspp}) with $\delta(\textbf{s})$ as in (\ref{delta1}),(\ref{pipa}) and (\ref{pppa})) \end{tabular}\\
    %6     & PA + PO & \begin{tabular}{@{}c@{}}Integration of PA and PO accounting for sampling and reporting process of CS: Option 2 \\(Equations (\ref{obspp}) with $\delta(\textbf{s})$ as in (\ref{delta2}),(\ref{pipa}) and (\ref{pppa})) \end{tabular} \\
    \bottomrule
    \end{tabular}%
  \label{modcomparison}%
\end{table}%

\subsection{Results}

\begin{figure}[H]
	\center
	\includegraphics[width=1\textwidth]{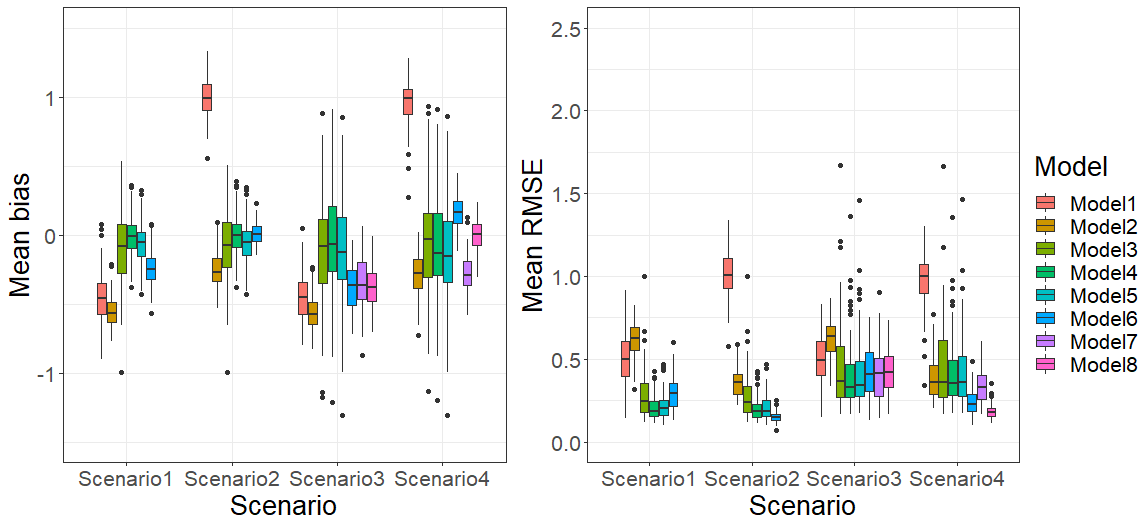} 
	\caption{Mean bias and RMSE for $\beta_1$ for each of the eight models proposed.}
	\label{fig:simu2}
\end{figure}

The results of this simulation study are summarized for each of the parameters of the model using bias and RMSE. %The intercept of the linear predictor is relevant when predicting the risk of powerline-induced death of birds as it determines the level of the predicted probabilities. Therefore, we compare the models through the aforementioned comparison measures in Figure \ref{fig:simu1}.\\ \\
%	\begin{figure}[H]
%	\center
%	\includegraphics[width=1\textwidth]{Figures/POWSimusBeta0v2409.png} 
%	\caption{Mean bias and RMSE for $\beta_0$ for each of the eight models proposed.}
%	\label{fig:simu1}
%\end{figure}
%Note that in presence of preferential sampling and high reporting probability (Scenario 2), model 6 is not the one that performs the best. It means that model 6 is not as efficient as model 5 to remove noise caused by the different biases in the collection of CS data from the intercept in Scenario  2.  Something similar occurs in Scenario 4, where models 2 (which uses only data from professional surveys), 7 and 8 (which account for some of the noise caused by low reporting effort) outperform model 6. In Scenario 3, as the probability of reporting a detected occurrence is low, CS data is yet noisy since uneven reporting effort has not been accounted for. Therefore, models 2,7 and 8 end up performing better than the other competing models. Scenarios 1 and 2 show the opposite, as reporting probability of a detected occurrence is quite high, model 5 (which accounts for both uneven sampling effort and detectability using only CS data) is the one that performs the best. Finally, the necessity of accounting for biases in the collection of both data types is evident as the estimate of the intercept for models 1 and 3 (which do not account for any source of bias) is much more biased than for the other models.\\ \\
The effect of the covariates on the ecological state is of paramount importance for use as decision support. Hence, in Figure \ref{fig:simu2} we present the performance measures for the parameter $\beta_1$, which in our simulation study represents the effect of the covariate \textit{cloud cover} on the risk of powerline-induced deaths.
 We first observe that model 1 (based only on professional surveys and not accounting for preferential sampling) performs poorly in scenarios with preferential sampling (scenarios 2 and 4), but outperforms model 2 (based only on professional surveys and accounting for preferential sampling) in scenarios with random sampling (Scenarios 1 and 3). Nevertheless, these two models were outperformed by all the models that are based on CS data (models 3-8) in all the simulation scenarios. Models based only on CS data (models 3, 4 and 5) performed better in scenarios with random selection of the sampled powerlines (scenarios 1 and 3). On the other hand, in the simulation scenario that assumed preferential selection of the powerlines and high willingness to report (Scenario 2), model 6 (both data types while accounting for preferential sampling and biases in CS data collection except reporting effort) performs much better than the other competing models showing, in addition, consistently less variability in the posterior means of the parameter $\beta_1$. Model 8 (both data types while accounting for preferential sampling and all biases in CS data collection) performed better than all the competing models in both bias and RMSE for the scenario with preferential sampling and low willingness to report (scenario 4). In this scenario model 6 outperformed model 7 (both data types and simple method to account for reporting effort) arguably due to the lack of structure of model 7 in the explanation of the differences in reporting effort. Finally, note that the variability in the posterior means of $\beta_1$ is much smaller in scenarios with high willingness to report (scenarios 1 and 2) for all the models, compared to scenarios with low willingness to report (scenarios 3 and 4) as much more information is available in CS data.\\ \\
 The summaries for both the fixed effects and the spatial hyperparameters are presented in Table S1 of the Supplementary Information. For $\beta_0$ we find out that models that do not account for any bias in the collection process (models 1 and 3) underestimate its value for all the simulation scenarios, while model 5 (based only on CS data and accounting for accessibility and detectability) is the one that performs the best for scenarios 1 and 2. Besides,  in scenarios with high willingness to report (scenarios 3 and 4), models 6, 7 and 8 (models that fusion both data types) outperformed all the other competing models. Note that the estimates of the spatial hyperparameters of the GRF $\omega_1(\textbf{s})$ are inaccurate for the most complex models (models 2, 6, 7 and 8). To explore in more detail the effect of this bias, we have fixed the spatial hyperparameters for one of the datasets of the simulation studies and have fitted models 6, 7 and 8. The marginal posterior distribution of $\beta_1$ in scenarios 3 and 4 is displayed in Figures S.11  and S.12 in the Supplementary Information. We arrive to similar results regarding the accuracy of the marginal posterior distributions for $\beta_1$. However, we noticed that these posterior distributions are more precise than when the spatial hyperaparameters were not fixed. Given the sensitivity of the posterior distributions of the spatial hyperparameters to the prior specification of the spatial hyperparameters in models 6, 7 and 8, more informative prior information could contribute to obtain more accurate posterior distributions for both the fixed effects and the spatial hyperparameters. \\ \\
Our proposed framework does not only allow us to infer the posterior distribution of the parameters that drive the ecological process, but also to learn about the processes that drive the biases in CS data and professsional surveys. In Tables S.1.-S.4. in the supplementary information, we present the comparison measures for the parameters involved in the thinning of the true point pattern to produce both CS and professional surveys data. We find out that the parameters of the sampling process model are similarly estimated by models 4 to 8, with small biases for both $\alpha_0$ and $\alpha_1$. On the other hand, the parameters associated with the detectability exhibit biases and large RMSE. This might be linked to poor identifiability of these parameters, which could be remediated with informative prior information or more information about the relation between land use and detectability. Finally, the parameters of the preferential sampling model are accurately estimated in Scenarios 2 and 4 as expected. A particularity of model 8 is that it allows for posterior inference about the willingness to report occurrences of powerline-induced deaths for each observer.\\ \\
Finally, we compare the proposed models in terms of predictive performance. We focus our comparison on two aspects: the accuracy of the predictions and how uncertain these predictions are. These predictions are only based on the fixed effects of the linear predictor. To compare the accuracy of the predictions produced, we computed the Root Mean Squared Error for the predicted probabilities by each model in each simulation scenario on a dense grid along the powerline network of Trøndelag. The RMSE maps for each model in each scenario are presented in Figures S.2.,S.4,S.6 and S.8. The results show that in scenarios with high willingness to report (scenarios 1 and 2), model 5 (based only on CS data and accounting for accessibility and detectability) outperforms all the other models. In scenarios with low willingness to report (scenarios 3 and 4), model 5 was outperformed by models that fusion both data types (models 6-8) in vast portions of the study region.\\ \\
Regarding the uncertainty of the predictions produced in the different simulation scenarios, we computed the average width of the 95\% prediction intervals for each model over the dense grid mentioned above. The maps with  the length of the prediction intervals are displayed in Figures S.3., S.5, S.7 and S.9. Model 2 (using only professional surveys while accounting for preferential sampling) produces the larger uncertainties, while the most complex models (models 6 to 8) have larger uncertainties than the models based only on CS data (models 3-5).
%Tasks: We clearly need the 95% PI width for XB, do it today!
% Tasks: Based on that change the speech

\section{Case study of bird mortality  and power lines in Trøndelag, Norway}

\begin{figure}[H]
	\center
	\includegraphics[width=1\textwidth]{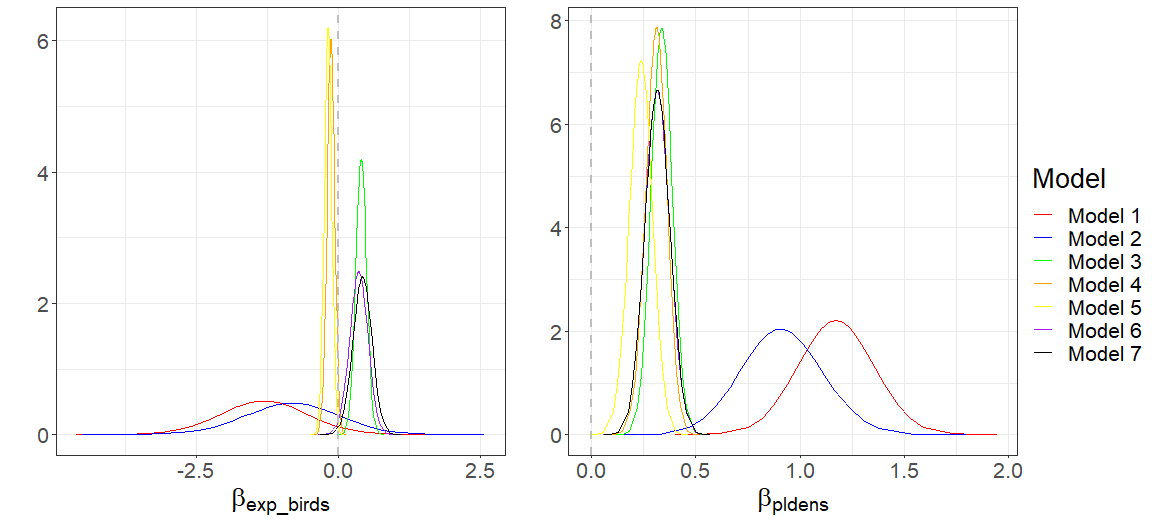} 
	\caption{Posterior distributions for the fixed effects of the seven models fitted}
	\label{fig:app1}
\end{figure}

\begin{figure}[H]
	\center
	\includegraphics[width=1\textwidth]{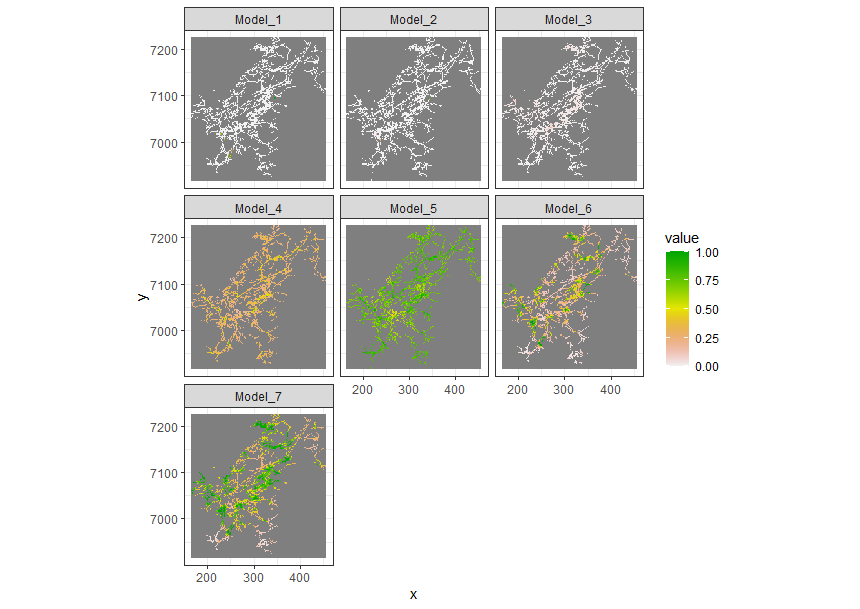} 
	\caption{Posterior median of the risk of powerline-induced deaths}
	\label{fig:app2}
\end{figure}

\begin{figure}[H]
	\center
	\includegraphics[width=1\textwidth]{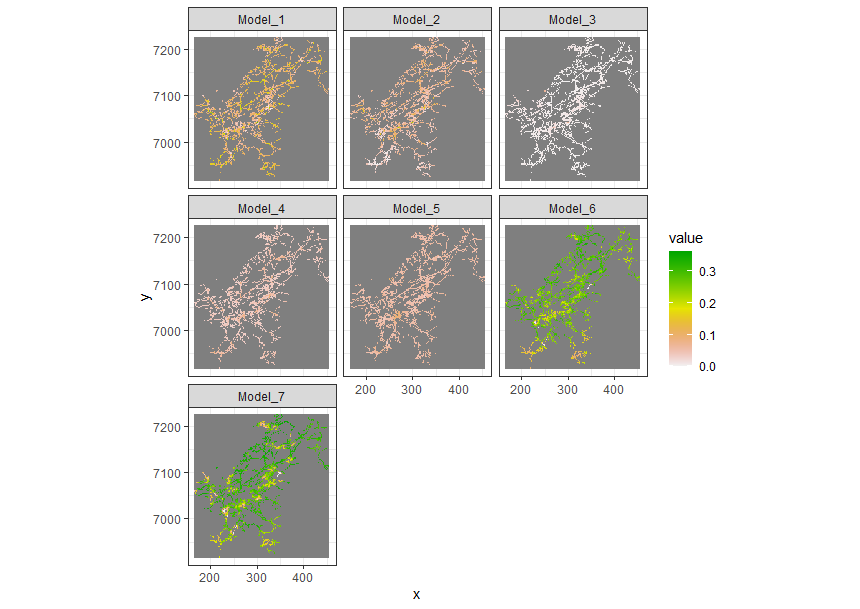} 
	\caption{Standard deviation of the predicted risk of powerline-induced deaths}
	\label{fig:app3}
\end{figure}

Our real data application study aims to %identify those locations in Trøndelag County (Norway) where the risk of collision or electrocution of birds due to powerlines is high and to ultimately produce maps of mortality hotspots to inform electricity companies and the Water and Energy Directorate (NVE), as well as to 
gain a better understanding of the role of the landscape (environmental covariates) in creating riskier regions for powerline-induced deaths for birds. Based on the effect of these on the risk of powerline-induced deaths, risk maps of powerline-induced deaths can be made to inform electricity companies and the Water and Energy Directorate (NVE). To achieve this goal we have two sources of information (see Section 2): 1) professional surveys performed by NINA and 2) opportunistic records collected by citizen scientists. In this section we fit models 1 to 7, presented in Section 4, then show model predictions on the risk of powerline-induced death of birds, and conclude by comparing the predictions obtained with the proposed models.\\ \\
The selection of the candidate variables to explain the processes behind both observed data types was based on expert knowledge and the exploratory analysis performed in Section 2. As higher occurrences of powerline-induced deaths are expected as more birds are exposed and the powerline network is dense, we have considered the covariates \textit{powerline density} (The Norwegian Water Resources and Energy Directorate;
NVE) and \textit{bird abundance} \citep{sicacha2022spatial} as the first candidate variables. \textit{Land use} (AR50, https://www.nibio.no/tema/jord/arealressurser/ar50), which explains the type of land around a powerline, and \textit{cloud cover} ((https://www.earthenv.org/cloud), which proxies the visibility the birds have as they fly, were also considered as candidate variables, but were discarded as land use was correlated with the two chosen covariates and adding cloud cover did not improved the existing models. As argued previously in \citet{fithian2015bias} and \citet{Monsarrat2019}, accessibility is one of the main factors that determine where citizen scientists are more prone to collect information on biodiversity. For this reason, the covariates \textit{distance to (tertiary) roads} and \textit{distance to water bodies} (sea, lakes and rivers) were considered to explain where citizen scientists collect their observations. Land use was not considered to explain where citizen scientists go to collect observations as this covariate is also correlated with the distance variables. Another factor that needs consideration is the detectability of a dead bird. This can be affected by the size of the bird \citep{borner2017bird,ponce2010carcass}, the time of the year the observation is made \citep{Bevanger1995}, the land use of the place where the carcass is located \citep{philibert1993counting,schutgens2014estimating,dominguez2020factors} among other factors. In our case, we consider the different land uses as proxies for explaining differences in detectability. Finally, whether or not citizen scientists are willing to report a dead bird is an unsolved question. Hence, we have opted for considering this as another relevant factor that affects what is observed in CS databases and account for it using  Model 7.\\ \\
Preferential sampling was also considered as a possible flaw in the collection process of the data from professional surveys performed by NINA since the experts leading these projects have prior knowledge they might use when determining where to collect observations. The results of Section 2 suggest powerline density and cloud cover as possible drivers for preferential sampling. According to information provided by NINA, elevation gradient was also relevant for choosing which lines to visit. Given its correlation with cloud cover (Pearson correlation coefficient, $\rho=0.76$), the variables chosen to explain the preferential sampling were the elevation gradient and powerline density.\\ \\
In Section 3, we pointed out the potential identifiability issues that may arise when trying to account for biases in CS data as the number of parameters to estimate increases considerably. Therefore, in addition to the two observed data types, we have also included a point pattern with the locations of CS reports so that the parameters of the sampling process can be identified. For the parameters that link the different land uses and the detection of dead birds, we have proposed informative prior distributions based on expert knowledge \citep{dominguez2020factors} as no studies that explain the relation between land use and detectability for Norway are available.
We fitted the seven proposed models while integrating the available data types for each source of bias in the collection process. The posterior summaries for each of the fixed effects involved in the ecological process are graphically presented in Figure \ref{fig:app1}. 
For the effect of the covariate \textit{Exposed birds} we observe considerable disagreement between the models proposed. Models 1,2, 4 and 5 have marginal posterior distributions centered below 0, whereas for models 3,6 and 7 this distribution is located mostly above 0. The effect of the covariate \textit{powerline density} is more consistent for all the models proposed, with marginal posterior distributions centered around the same values, except for models 1 and 2. It is worth noting that the posterior distributions of models 1 and 2 show much higher uncertainty than for the other models. The posterior summaries for these and the other parameters in the model are presented in the supplementary information. Note that, as in the simulation study, the spatial range parameter and the marginal variance took much higher values for models 2, 6 and 7 that for models 3, 4 and 5.\\ \\
The risk of powerline-induced deaths was predicted along the powerline network in Trøndelag using each of the seven models proposed. The posterior medians of the risk are presented in Figure \ref{fig:app2}
As seen in Table S.5., the posterior means of the intercept in models 1, 2 and 3 make the posterior median of the predicted risk to take low values, except for locations where sampling has been performed. More realistic predictions are produced by models 4-7. Model 4 highlights very few spots in the study region, while model 5 predicts larger risk all over the region after accounting for differences in detectability. Once both data types have been integrated (models 6 and 7) fewer areas are highlighted as riskier for powerline-induced deaths. Although similar, the probabilities predicted using model 7 are higher than using model 6 as the former accounts for differences in reporting effort. The uncertainty of the predictions was computed through the standard deviation of the predicted risks and is presented in Figure \ref{fig:app3}.
As expected, the predictions based only on professional surveys, which cover a small portion of the study area, have higher uncertainty than those obtained using only opportunistic records in CS data. As seen in Figure S.10., the uncertainty of the term $\omega_1(\textbf{s})$ is considerable for models 6 and 7 across all the study area, while the uncertainty of the fixed effects is larger for models 1 and 2. After transforming to the scale of the risk of powerline-induced death, models 6 and 7 are the most uncertain.\\ \\
These results are consistent with the predictions made for the simulated datasets in Section 4 as models that use only CS data produce the less uncertain predictions. However, models 6 and 7, which use only data from professional surveys, produce predictions with higher uncertainty than all the other models.

\section{Discussion}

In this paper we have proposed and evaluated a methodological framework for integrating multiple data types to address questions in ecology and biodiversity. We focus on two types of data to integrate: professional surveys and opportunistic records collected by citizen scientists. A fundamental assumption of this work is that the observations in both data types have the same underlying ecological process as their origin, but they have different observation and reporting processes, which generates observations with different spatial coverages and different observation efforts. We approached the challenge of integrating these data types in first place by modeling the factors that determine the generation of each data type. %figuring out how each data type was generated and which factors could potentially affect their collection process. 
In particular, the data collected by experts is affected by prior knowledge of the professionals that conduct these surveys, thus a preferential sampling design that gives some powerlines higher chances of being visited than others is assumed. On the other hand, CS data is affected by factors accessibility, detectability and uneven reporting effort among citizen scientists. Both data types were managed as realizations of thinned point processes, but with different processes behind the thinning of the actual ecological process. While CS data was assumed affected by processes that occur at point level, we assumed that data from professional surveys was affected by selection probabilities occurring at areal level.\\ \\
To account for the factors that affected the collection process of both data types and avoid identifiability issues \citep{fithian2015bias}, we used additional sources of information that provided knowledge of the processes involved in the thinning. As no previous studies that inform about the link between land use/habitat and detectability of dead birds was available for our study region, priors based on expert knowledge were proposed. We proposed Bayesian spatial models that resemble the generation of each data type and lie within the class of Latent Gaussian Models, hence they were fitted using the INLA-SPDE approach as this is a computationally efficient approach for fitting this class of spatial Bayesian models. However, as seen in Section 3, the terms of the linear predictor that explain the effect of biases in the collection process on the generation of the observed point patterns do not follow the log-linear requirement underlying the INLA methodology for log-Gaussian Cox processes. For this reason, as a complement to the traditional INLA-SPDE approach, we used the approach in the \textit{inlabru} R-package \citep{bachl2019inlabru}, which is based on an iterative linearization of the non-linear terms in the linear predictor. This extra step might represent additional computational burden when a GRF is part of the nonlinear part of the linear predictor since the sparsity induced by the SPDE approach and the approximation of the Gaussian Random Fields as Gaussian Markov Random Fields might be reduced. For the most complex models (models 7 and 8 defined in Section 4), we experienced some numerical issues that might be related to this approximation as convergence was not reached.\\ \\
A simulation study and a case study were performed in order to study characteristics and properties of the models proposed in Section 3. Models that used one or two data types and that accounted for biases in the collection process to different extents were fitted for 100 different datasets in scenarios with professional surveys that select powerlines preferentially and randomly and with high and low willingness to report powerline-induced deaths by citizen scientists. The results of the simulation studies show the relevance of integrating both data types in scenarios with preferential sampling as model 6-8 produce more accurate estimates for the fixed effects in Scenarios 2 and 4 (scenarios with preferential sampling),  which are the scenarios that more closely resemble what occurs in the case study. In Scenarios 1 and 3 (scenarios without preferential sampling), models that account for preferential sampling do not perform as well as in the other scenarios, probably due to lack of identifiability of the large amount of parameters these models introduce.\\ \\
The assessment of the predictive performance of the proposed models show that models that integrate both data types (models 6-8) performed better than those based on only one data type (models 1-5). Regarding the uncertainty of the predictions, those models based only on professional surveys (models 1 and 2) predicted with larger uncertainty than those models based on only CS data (models 3-5), likely due to the larger area covered by CS data. Also the models that integrated both data types (models 6-8) produced predictions with larger uncertainty than model based only on CS data (models 3-5).
%Regarding the predictive performance of the models, in the simulation studies models 1-3 produced inaccurate predictions, while models 5-8 performed similarly, with model 5 producing the most accurate predictions. In our simulation studies models 1 and 2 produced predictions with larger uncertainty as they make use of only one data type which covers very small portions of the space. The uncertainty in models with only data from professional surveys affects as well those models that integrate both data types because their predictions are still more uncertain than those models that use only CS data.
A similar pattern was observed in our case study, where predictions using models 1, 2, 6 and 7 had higher uncertainty than the predictions of models 3-5.\\ \\ 
In the simulation studies we found that the posterior estimates of the spatial hyperparameters of the most complex models (models 2 and 6-8) were considerably more biased compared to the estimates obtained using only CS data. In particular, in both the simulation studies and the case study the spatial range of $\omega_1(\textbf{s})$ was much larger for these models. Arguably the choice of prior distribution is more influential on the models based only on professional surveys (models 1 and 2) as the spatial coverage of the professional surveys is small.\\ \\
Through the simulation studies we noticed that the proposed framework contributes to account for both the fixed effects behind an ecological process and the spatial autocorrelation that determines it. Besides, this framework has been proven useful as well to account for and quantify the factors that affect the collection process of both CS data and professional surveys. Hence, the estimates of our models can be used to device more informed sampling of CS data by focusing sampling efforts in areas with higher uncertainty or with low sampling effort. The simulation studies also showed the importance of knowing more about the preferences of citizen scientists. If more is known about these preferences, more targeted activities for citizen scientists can be launched in order to target specific research questions and the differences in reporting effort could be accounted for better inference. \\ \\
%The observed predictive performance of the most complex models (models 2, 6-8) is arguably related to the lack of convergence of the iterative INLA procedure used to linearize part of the linear predictor. More efforts are needed to achieve faster convergence and hence avoid issues with the estimates of the spatial hyperparameters that can potentially end up in more inaccurate prediction. Alternatively, a sensitivity analysis on models 2,6,7 and 8 could provide more insight into how to deal with the biased estimates of the hyperparameters of the Gaussian Random Field $\omega_1(\textbf{s})$. The simulation studies also showed the importance of knowing more about the preferences of citizen scientists. If more is known about these preferences, more targeted activities for citizen scientists can be launched in order to target specific research questions and the differences in reporting effort could be accounted for better.\\ \\
The case study of deaths caused by powerlines in Trøndelag, Norway motivated this
%addresses the research question that motivated the methods developed in this 
paper. Through this case study we showed the importance of proposing methods that combine more than one data type as we showed how the effect of a factor like the amount of exposed birds varies amongst models. Moreover, the prediction maps produced highlighted zones with higher risk of powerline-induced deaths. These maps could be used by conservation programs to target mitigation measures for powerlines with higher risks.\\ \\
%This modeling framework relies on the INLA-SPDE approach and a linearization step available in the \textit{inlabru} R-package. This additional step allows to explicitly quantify for biases in the collection process of both data types, which has allowed to perform extensive simulation studies. That said, the extra step represents additional computational burden when a GRF is part of the nonlinear part of the linear predictor since the sparsity induced by the SPDE approach and the approximation of the Gaussian Random Fields as Gaussian Markov Random Fields might be reduced. Moreover, convergence was not reached for models 7 and 8 when the \textit{inlabru} approach was used.\\ \\
%The uncertainty quantified by this modeling framework as well as the quantification of the effect of different types of bias in the CS data collection process can serve as a first step in the design of new surveys by focusing sampling efforts in areas with higher uncertainty or with low sampling effort. 
The proposed methods in this paper are made available also through code such that practitioners in ecology and biodiversity can use them to address many other questions using multiple sources of information while accounting for other sources of bias given the flexible specification of the models for both professional surveys and CS data.

	\section*{Bibliography}
	\begingroup
	\renewcommand{\section}[2]{}%
	\bibliography{POWRefsF}
	\endgroup
%% ** Mainmatter **

%\section{}\label{}

% \begin{figure} 
% \includegraphics{<eps-file>}% place <eps-file> in ./img  subfolder
% \caption{}
% \label{}
% \end{figure}

% \begin{table} 
% *****************
% \begin{tabular}{lll}
% \end{tabular}
% *****************
% \caption{}
% \label{}
% \end{figure}

%%%%%%%%%%%%%%%%%%%%%%%%%%%%%%%%%%%%%%%%%%%%%%
%% Supplementary Material, if any, should   %%
%% be provided in {supplement} environment  %%
%% with title and short description.        %%
%%%%%%%%%%%%%%%%%%%%%%%%%%%%%%%%%%%%%%%%%%%%%%
%\begin{supplement}
%\stitle{???}
%\sdescription{???.}
%\end{supplement}

%% ** The bibliograhy **
%\bibliographystyle{ba}
%\bibliography{<bib-data-file>}% place <bib-data-file> in ./bib folder 

% ** Acknowledgements **
%\begin{acks}[Acknowledgments]
%\end{acks}

\end{document}